# Effective electronic-only Kohn-Sham equations for the muonic molecules

Milad Rayka[1], Mohammad Goli[2,*] and Shant Shahbazian[1,*]

[1] *Department of Physics and Department of Physical and Computational Chemistry, Shahid Beheshti University, G. C., Evin, Tehran, Iran, 19839, P.O. Box 19395-4716.*

[2] *School of Nano Science, Institute for Research in Fundamental Sciences (IPM), Tehran 19395-5531, Iran*

E-mails:

Mohammad Goli : mgoli_chem@yahoo.com

Shant Shahbazian: sh_shahbazian@sbu.ac.ir

[*] Corresponding authors

## Abstract

A set of effective electronic-only Kohn-Sham (EKS) equations are derived for the muonic molecules (containing a positively charged muon), which are completely equivalent to the coupled electronic-muonic Kohn-Sham equations derived previously within the framework of the Nuclear-Electronic Orbital density functional theory (NEO-DFT). The EKS equations contain effective non-coulombic external potentials depending on parameters describing muon's vibration, which are optimized during the solution of the EKS equations making muon's KS orbital reproducible. It is demonstrated that the EKS equations are derivable from a certain class of effective electronic Hamiltonians through applying the usual Hohenberg-Kohn theorems revealing a "duality" between the NEO-DFT and the effective electronic-only DFT methodologies. The EKS equations are computationally applied to a small set of muoniated organic radicals and it is demonstrated that a mean effective potential maybe derived for this class of muonic species while an electronic basis set is also designed for the muon. These computational ingredients are then applied to muoniated ferrocenyl radicals, which had been previously detected experimentally through adding muonium atom to ferrocene. In line with previous computational studies, from the six possible species the staggered conformer, where the muon is attached to the exo position of the cyclopentadienyl ring, is deduced to be the most stable ferrocenyl radical.

# I. Introduction

The bound states of the muonic molecules, derived from the attachment of the positively charged muon, $\mu^+$, to usual molecules, have been in focus in recent years through the $\mu^+$ spin rotational/relaxation/resonance (μSR) spectroscopy with vast applications from condensed matter physics to chemistry and molecular biology [1-8]. The basic measured quantities in the μSR spectroscopy are the hyperfine coupling constants which are used to locate the molecular site where the $\mu^+$ is attached [1,3]. However, the assignment procedure of the hyperfine coupling constants is not always an easy task since in complex molecules there are several sites potentially capable of trapping $\mu^+$. Thus, it is desirable to derive theoretically both preferred sites for $\mu^+$ addition and their corresponding hyperfine coupling constants from quantum mechanical calculations [9-28]. The commonly-used molecular quantum mechanical methods are based on the clamped nucleus paradigm conceiving electrons as quantum particles and the nuclei as point charges [29,30]. Nevertheless, it is not generally evident that to what extent a light-mass $\mu^+$, which is virtually one-ninth of proton's mass, could be properly treated as a point charge [22-28]. One possible response to this concern is treating $\mu^+$ as a quantum particle like electron and trying to incorporate simultaneously the kinetic energy operators of electrons and $\mu^+$ into the Hamiltonian of the muonic molecules for quantum mechanical calculations. Recently, we have employed this strategy to consider several muonic molecules within the context of the Nuclear-Electronic Orbital (NEO) ab initio methodology [31-37]. As expected, one may extend such studies to more complex muonic molecules and try to develop more accurate ab initio computational procedures to achieve experimental accuracy. Nevertheless, the recent "effective" reformulation of the NEO-Hartree-Fock (NEO-HF) equations [38,39], based on a simplified wavefunction proposed by Auer and Hammes-Schiffer [40], opens a new door to design "muon-specific" ab initio procedures. The present study is also a continuation of the same path trying to incorporate electron-electron (ee) correlation, which is absent in the effective NEO-HF (EHF) method, into the effective NEO theory. For this purpose, one may pursue two independent paths; trying to incorporate the ee correlation using various wavefunction-based post-NEO-HF procedures or using the NEO-density functional theory (NEO-DFT). In this report, the latter possibility is considered through introducing effective "electronic-only" Kohn-Sham (KS) equations for the muonic molecules while the former option will be considered in a subsequent report.

The idea of extending DFT to systems containing multiple quantum particles, i.e., electrons plus at least one other type of quantum particles, is not new and have been tried for the electron-hole and the electron-positron systems many decades ago [41-43]. However, the seminal paper by Parr and coworkers is usually perceived as the first rigorous formulation of the multi-component DFT [44], though since then various extensions have been proposed [45-49]. Particularly, the recent interest in developing orbital-based ab initio non-Born-Oppenheimer procedures treating both electrons and nuclei as quantum particles from outset [31-33,50-53], triggered a renewed interest in practical reformulation and computational implementation of the multi-component DFT. After the pioneering studies [48,54], a number of papers have appeared dealing with various aspects of computational implementation and the functional design of the multi-component DFT in non-Born-Oppenheimer realm [55-69]. The NEO-DFT as proposed and extended by Hammes-Schiffer and coworkers in recent years belongs to this category [56,62-68]. In the NEO-DFT, in contrast to the usual electronic DFT [70-73], one is faced with the coupled KS equations for each type of quantum particles, and therefore, each equation contains its own effective KS potential energy with some unique terms. These terms appear from multitude of the exchange-correlation functionals and while the previously designed electronic exchange-correlation functionals are used to describe the ee correlation, new functionals for other types of quantum particles and their interactions must be designed. In the case of the muonic molecules the basic types of correlations are the ee and eµ correlations and based on these correlations the muonic NEO-DFT may be divided into NEO-DFT(ee) and NEO-DFT(ee+eµ) categories. The latter theory is, in principle, exact though needs the proper introduction of the eµ correlation functional while the former theory basically neglects the eµ correlation. For systems containing quantum protons there have been several attempts to introduce electron-proton correlation functionals [48,49,59-67]. However, no systematic comparative study has been conducted yet to compare the relative merits of the proposed functionals though many (but not all) of them seem to be inspired in some way from the original Colle-Salvetti formula for the ee correlation [74]. In this study, we will neglect the eµ correlation and an effective formulation of the NEO-DFT(ee) is proposed, similar to that proposed for the NEO-HF theory [38,39].

The paper is organized as follows. In section II the effective NEO-DFT(ee) and corresponding KS equations are discussed while our computational details are provided in Section III. In section IV, the optimized electronic basis sets used for $\mu^+$ are introduced through considering a small but representative set of organic molecules. In addition, the computational implementation of the effective equations for large molecules with an example is studied to demonstrate the efficiency of the proposed theory. Finally, our conclusions are offered in Section V.

## II. Theory

The Hamiltonian for a system containing $N$ electrons, a single massive quantum positively charged quantum particle (PCP) with mass $m$ (assuming $m \gg m_e$) and $q$ clamped nuclei, is as the following in the atomic units ($m_e = \hbar = 1$):

$$\hat{H}_{total} = \hat{H}_{NEO} + \sum_{\beta}^{q}\sum_{\gamma\rangle\beta}^{q} \frac{Z_\beta Z_\gamma}{|\vec{R}_\beta - \vec{R}_\gamma|}$$

$$\hat{H}_{NEO} = (-1/2)\sum_i^N \nabla_i^2 + (-1/2m)\nabla_\mu^2 - \sum_i^N \frac{1}{|\vec{r}_\mu - \vec{r}_i|} + \sum_i^N \sum_{j\rangle i}^N \frac{1}{|\vec{r}_i - \vec{r}_j|} + V_{ext}$$

$$V_{ext} = -\sum_i^N \sum_\beta^q \frac{Z_\beta}{|\vec{R}_\beta - \vec{r}_i|} + \sum_\beta^q \frac{Z_\beta}{|\vec{R}_\beta - \vec{r}_\mu|} \qquad (1)$$

This is a two-component NEO Hamiltonian that includes the basic physics of the muonic molecules if the mass of the PCP to be fixed at the mass of $\mu^+$ ($m = 206.768 m_e$). Since the formalism of the NEO-DFT has been extensively discussed for the general multi-component systems, herein, only the main points are discussed briefly and the interested reader may consult the original literature for details [56,62-67]. In principle, not only $\mu^+$ but also all nuclei may be treated as quantum particles and a DFT formalism for resulting "self-bound" muonic system can be devised [75-78], however, this generalized formalism is beyond the scope of this paper. It is possible to demonstrate that the external potential, $V_{ext}$, is uniquely determined by the following one-particle densities:

$$\rho_e(\vec{r}) = N\int d\vec{r}_2 ... \int d\vec{r}_N \int d\vec{r}_\mu \sum_{spins} \Psi^*\Psi \ , \qquad \rho_\mu(\vec{r}_\mu) = \int d\vec{r}_1 ... \int d\vec{r}_N \sum_{spins} \Psi^*\Psi \quad (2)$$

In these equations $\Psi$ and its complex conjugate stand for the muonic ground state wavefunction while the resulting universal functional in the Levy's constrained search procedure depends on both densities and the mass of the PCP (for a very brief but lucid discussion see section 9.6 in [72]). Assuming the KS reference system to be a non-interacting system of electrons and the PCP the KS wavefunction is a product of a determinant composed of the KS spin-orbitals for electrons and a single spatial orbital for the PCP. The total energy of the system for a typical closed-shell electronic system, neglecting eµ correlation, is the following:

$$E_{NEO} = 2\sum_{i}^{N/2}\int d\vec{r}_i\, \psi_i^*(\vec{r}_i)\hat{h}_i\psi_i(\vec{r}_i) + \int d\vec{r}_\mu\, \psi_\mu(\vec{r}_\mu)\hat{h}_\mu\psi_\mu(\vec{r}_\mu)$$

$$+J_{ee}[\rho_e] + E_{exc}[\rho_e] - \int d\vec{r}\int d\vec{r}_\mu\, \frac{\rho_e(\vec{r})\rho_\mu(\vec{r}_\mu)}{|\vec{r}-\vec{r}_\mu|}$$

$$\hat{h}_i = (-1/2)\nabla_i^2 - \sum_{\beta}^{q}\frac{Z_\beta}{|\vec{R}_\beta - \vec{r}_i|},\quad \hat{h}_\mu = (-1/2m)\nabla_\mu^2 + \sum_{\beta}^{q}\frac{Z_\beta}{|\vec{R}_\beta - \vec{r}_\mu|}$$

$$J_{ee}[\rho_e] = \int d\vec{r}_1\int d\vec{r}_2\, \frac{\rho_e(\vec{r}_1)\rho_e(\vec{r}_2)}{|\vec{r}_1 - \vec{r}_2|} \qquad (3)$$

In the preceding equations $E_{exc}$, the exchange-correlation energy functional, stands for all so-called "non-classical" effects resulting from the electron exchange, ee correlation and residual electronic kinetic energy beyond the KS reference system [71-73]. If the energy is varied with respect to the electronic and muonic spatial orbitals the following coupled KS equations are derived:

$$\left[(-1/2)\nabla_1^2 + v_{KS}^e(\vec{r}_1)\right]\psi_j(\vec{r}_1) = \varepsilon_j\psi_j(\vec{r}_1),\quad j=1,...,N/2$$

$$\left[(-1/2m)\nabla_\mu^2 + v_{KS}^\mu(\vec{r}_\mu)\right]\psi_\mu(\vec{r}_\mu) = \bar{\varepsilon}_\mu\psi_\mu(\vec{r}_\mu)$$

$$v_{KS}^{e}(\vec{r}_{1}) = -\sum_{\beta}^{q} \frac{Z_{\beta}}{|\vec{R}_{\beta} - \vec{r}_{1}|} + \int d\vec{r} \frac{\rho_{e}(\vec{r})}{|\vec{r} - \vec{r}_{1}|} - \int d\vec{r}_{\mu} \frac{\rho_{\mu}(\vec{r}_{\mu})}{|\vec{r}_{\mu} - \vec{r}_{1}|} + v_{exc}(\vec{r}_{1})$$

$$v_{KS}^{\mu}(\vec{r}_{\mu}) = \sum_{\beta}^{q} \frac{Z_{\beta}}{|\vec{R}_{\beta} - \vec{r}_{\mu}|} - \int d\vec{r} \frac{\rho_{e}(\vec{r})}{|\vec{r} - \vec{r}_{\mu}|}$$

$$v_{exc} = \frac{\delta E_{exc}[\rho_{e}]}{\delta \rho_{e}} \qquad (4)$$

These differential equations are usually transformed into algebraic equations by expanding all orbitals through known basis sets and then solving algebraic equations employing the self-consistent field procedure (extension to open-shell electronic systems is also straightforward and is not considered herein).

As discussed in the previous communications [38,39], since the PCP is well-localized due to its large mass relative to that of electron, the orbital of the PCP may be approximated with a wavefunction describing a 3D quantum oscillator. The simplest example is an isotropic harmonic oscillator with the following ground state wavefunction: $\psi_{\mu}(\vec{r}_{\mu}) = (2\alpha/\pi)^{\frac{3}{4}} Exp(-\alpha|\vec{r}_{\mu} - \vec{R}_{c}|^{2})$, where $\alpha$ is the width and $\vec{R}_{c}$ is the center of the Gaussian-type function (GTF), which are the standard parameters of a quantum oscillator [38]. It is possible to use more complicated anharmonic and anisotropic oscillator models with a large set of parameters instead, as discussed in detail elsewhere [39], however, the isotropic harmonic oscillator may be employed as an illustrative example of the formulation of the effective NEO-DFT(ee) (for a similar idea see [79]). Incorporating this s-type GTF in equation (3) yields the following expression for the energy:

$$E_{NEO} = 2\sum_{i}^{N/2} \int d\vec{r}_{i}\, \psi_{i}^{*}(\vec{r}_{i}) \hat{h}_{i} \psi_{i}(\vec{r}_{i}) - \int d\vec{r} \frac{\rho_{e}(\vec{r})}{|\vec{r} - \vec{R}_{c}|} erf\left[\sqrt{2\alpha}|\vec{r} - \vec{R}_{c}|\right]$$

$$+ J_{ee}[\rho_e] + E_{exc}[\rho_e] + U$$

$$U(\alpha, \vec{R}_c, \{\vec{R}_\beta\}) = \left(\frac{3\alpha}{2m}\right) + \sum_\beta^q \frac{Z_\beta}{|\vec{R}_\beta - \vec{R}_c|} erf\left[\sqrt{2\alpha}\,|\vec{R}_\beta - \vec{R}_c|\right] \quad (5)$$

In this equation, the PCP disappears as a quantum particle and instead novel non-coulombic potential energy terms appear containing the parameters of the original GTF. Particularly, $U$ can be conceived as a classical potential energy term similar to the nuclear repulsion term in equation (1) and does not need to be considered in subsequent functional variation. Thus, the total and effective KS energies are now:

$$E_{total} = E_{eff-KS} + E_{classical}$$

$$E_{eff-KS} = 2\sum_i^{N/2} \int d\vec{r}_i\, \psi_i^*(\vec{r}_i)\hat{h}_i\psi_i(\vec{r}_i) - \int d\vec{r}\, \frac{\rho_e(\vec{r})}{|\vec{r}-\vec{R}_c|} erf\left[\sqrt{2\alpha}\,|\vec{r}-\vec{R}_c|\right]$$

$$+ J_{ee}[\rho_e] + E_{exc}[\rho_e]$$

$$E_{classical}(\alpha, \vec{R}_c, \{\vec{R}_\beta\}) = U + \sum_\beta^q \sum_{\gamma > \beta}^q \frac{Z_\beta Z_\gamma}{|\vec{R}_\beta - \vec{R}_\gamma|} \quad (6)$$

Upon variation of $E_{eff-KS}$ with respect to the electronic orbitals the following effective electronic-only KS equations, hereafter briefly called the EKS equations, arise:

$$\left[(-1/2)\nabla_1^2 + v_{eff-KS}^e(\vec{r}_1)\right]\psi_j(\vec{r}_1) = \varepsilon_j \psi_j(\vec{r}_1), \quad j=1,\ldots N/2$$

$$v_{eff-KS}^e(\vec{r}_1) = -\sum_\beta^q \frac{Z_\beta}{|\vec{R}_\beta - \vec{r}_1|} + v_{eff}^{e\mu}(\vec{r}_1) + \int d\vec{r}\, \frac{\rho_e(\vec{r})}{|\vec{r}-\vec{r}_1|} + v_{exc}(\vec{r}_1)$$

$$v_{eff}^{e\mu}(\vec{r}_1) = -\frac{1}{|\vec{R}_c - \vec{r}_1|} erf\left[\sqrt{2\alpha}\,|\vec{R}_c - \vec{r}_1|\right] \quad (7)$$

Formally, solving the EKS equations is equivalent to the solution of the coupled equations offered in equation (4) with a single GTF as a basis set. In practice the price that has been payed is adding two new parameters of the GTF, i.e. $\alpha$ and $\vec{R}_c$, in the optimization procedure of $E_{classical}$ along the usual geometry optimization of the clamped nuclei. Like the case of equations (4) these differential equations may be transformed into algebraic equations by expanding the electronic KS orbitals in GTFs. The one-electron integrals resulting from the electron-PCP interaction potential energy term, $v_{eff}^{e\mu}(\vec{r}_1)$, are available analytically [80]. Still, even in the absence of analytical formulas, the same numerical integration procedure employed to derive the integrals associated to $v_{exc}(\vec{r}_1)$ may also be used to evaluate this type of integrals [81-84]. At this stage of development, the algebraic EKS equations can be solved by using any known electronic exchange-correlation functionals and basis sets for the study of muonic system. However, before considering computational implementation, let us try to grasp certain ramifications of the effective formulation.

Although equations (7) were derived assuming the two-component Hamiltonian and associated NEO-DFT, it is possible to reverse the procedure and try to construct an effective electronic Hamiltonian that yields equations (7) directly through its own DFT (for a comprehensive discussion on "building" DFTs for "model" Hamiltonians see [85]). The following Hamiltonian is a proper candidate:

$$\hat{H}_{eff} = \hat{H}_{elec-eff} + \sum_{\beta}^{q}\sum_{\gamma\rangle\beta}^{q} \frac{Z_\beta Z_\gamma}{|\vec{R}_\beta - \vec{R}_\gamma|} + \frac{3\alpha}{2m} + \sum_{\beta}^{q} \frac{Z_\beta}{|\vec{R}_\beta - \vec{R}_c|} erf\left[\sqrt{2\alpha}\left|\vec{R}_\beta - \vec{R}_c\right|\right]$$

$$\hat{H}_{elec-eff} = (-1/2)\sum_{i}^{N}\nabla_i^2 + \sum_{i}^{N}\sum_{j\rangle i}^{N} \frac{1}{|\vec{r}_i - \vec{r}_j|} + V_{ext}$$

$$V_{ext} = -\sum_{i}^{N}\sum_{\beta}^{q} \frac{Z_\beta}{|\vec{R}_\beta - \vec{r}_i|} - \sum_{i}^{N} \frac{1}{|\vec{R}_c - \vec{r}_i|} erf\left[\sqrt{2\alpha}\left|\vec{R}_c - \vec{r}_i\right|\right] \quad (8)$$

It is straightforward to demonstrate that the whole machinery of the electronic DFT is equally applicable, and equations (7) arise assuming a non-interacting electronic KS reference system for the effective electronic Hamiltonian. It is interesting to try to generalize this result and introduce a picture not tied to specific vibrational models used for the PCP (for a discussion on complicated vibrational models beyond harmonic oscillator see [39]). Accordingly, there is a correspondence between the effective potential and the Gaussian or Slater type basis sets (or any other well-designed mathematical function) used to expand the orbital of the PCP. This correspondence originates from integrating the kinetic energy integral of the PCP and the electron-PCP interaction integrals which leads to the remaining of only basis function parameters, denoted as $\{c_k\}$. Thus, the resulting correspondence of the PCP orbital and the effective interaction potential is: $\psi_\mu(\vec{r}_\mu;\{c_k\}) \to v_{eff}^{e\mu}(\vec{r}_1;\{c_k\}), U(\{c_k\})$ [39]. Hence, the most general form of the EKS equations, reiterating equations (5) to (7), is:

$$\left[(-1/2)\nabla_1^2 + v_{eff-KS}^e(\vec{r}_1)\right]\psi_j(\vec{r}_1) = \varepsilon_j \psi_j(\vec{r}_1), \quad j = 1,\ldots N/2$$

$$v_{eff-KS}^e(\vec{r}_1) = -\sum_\beta^q \frac{Z_\beta}{|\vec{R}_\beta - \vec{r}_1|} + \int d\vec{r}\, \frac{\rho_e(\vec{r})}{|\vec{r} - \vec{r}_1|} + v_{eff}^{e\mu}(\vec{r}_1;\{c_k\}) + v_{exc}(\vec{r}_1)$$

$$E_{classical}(\{c_k\},\{\vec{R}_\beta\}) = U(\{c_k\}) + \sum_\beta^q \sum_{\gamma > \beta}^q \frac{Z_\beta Z_\gamma}{|\vec{R}_\beta - \vec{R}_\gamma|} \tag{9}$$

In these equations $\{c_k\}$ must be optimized like the nuclear geometry and using the optimized parameters, one may reconstruct the orbital of the PCP and corresponding one-particle density, which describes the vibrational motion of the PCP. The corresponding effective Hamiltonian may be written as the following:

$$\hat{H}_{eff} = \hat{H}_{elec-eff} + \sum_{\beta}^{q}\sum_{\gamma\rangle\beta}^{q} \frac{Z_{\beta}Z_{\gamma}}{\left|\vec{R}_{\beta}-\vec{R}_{\gamma}\right|} + U(\{c_k\})$$

$$\hat{H}_{elec-eff} = (-1/2)\sum_{i}^{N}\nabla_i^2 + \sum_{i}^{N}\sum_{j\rangle i}^{N}\frac{1}{\left|\vec{r}_i-\vec{r}_j\right|} + V_{ext}$$

$$V_{ext} = -\sum_{i}^{N}\sum_{\beta}^{q}\frac{Z_{\beta}}{\left|\vec{R}_{\beta}-\vec{r}_i\right|} + \sum_{i}^{N}v_{eff}^{e\mu}(\vec{r}_i;\{c_k\}) \tag{10}$$

Equations (9) and (10) are the heart of the effective NEO-DFT for the muonic systems though it can be used also for systems containing a single quantum proton or any other heavier particle and may easily be extended to the multi-component cases with more than two types of quantum particles as well.

## III. Computational details

Based on the proposed effective DFT in order to start solving the EKS equations, at first step, the effective potentials must be introduced. In this study the used potentials are constructed from a fully-optimized single s-type GTF, [1s], and from a scaled [5s5p] Gaussian basis set; the former has been explicitly given in equation (7). It must be emphasized that instead of deriving the potential separately for each basis set, an automated algorithm may be constructed to produce the potential after determining the type and number of GTFs used to expand the muonic orbital (Goli and Shahbazian, under preparation). For describing electronic distribution, 6-311++g(d,p) basis set was placed on the clamped nuclei [86-88], while for the muon a [4s1p] electronic basis set was placed at a *banquet* atom; in the [1s] associated effective potential, this center is denoted by $\vec{R}_c$ in equations (7). In all ab initio calculations, the B3LYP exchange-correlation hybrid functional was employed for electrons without further modifications or re-optimization of its parameters [89-91] (we leave the possibility of reparametrizing this functional for the muonic systems to a future study). The whole computational level is termed EKS-B3LYP/[6-311++g(d,p)/4s1p]. The optimized parameter of the effective potential, $\alpha$, in equations (7) as well as the energy-optimized exponents of [4s1p] electronic basis set of a representative set of organic molecules (*vide infra*) were determined through a full optimization of the EKS equations using a non-linear numerical optimization procedure [39]. In the case of [5s5p] associated effective potential partial optimization was done and only half of parameters, the linear coefficients (*vide infra*), were determined through direct optimization as discussed in the next section. Besides, the energy-optimized exponents of [4s1p] electronic basis derived from the [1s] associated effective potential were used without further optimization for the EKS calculations with [5s5p] associated effective potential. The geometry of the clamped and banquet nuclei was optimized using the analytical gradients during the geometry optimization procedure. Since all considered muonic species are odd-electron systems, the unrestricted (U) as well as the restricted-open (RO) versions of the algebraic EKS equations were utilized for ab initio calculations. A modified version of the GAMESS package was used for all ab initio calculations and the original implemented numerical integration algorithm for the exchange-correlation integrals was employed without any modifications [92,93]. Throughout the calculations, the used masses for $\mu^+$, proton (H), deuterium (D) and tritium (T) are 206.76828, 1836.15267, 3670.48296 and 5496.92153, respectively, in atomic units.

**IV. Results and discussion**

    **A. Designing the effective potential and the electronic basis set**

In the previous EHF study on a large set of species it was demonstrated that the parameters of the effective potential as well as the exponents of the electronic basis set are mainly determined by the mass of the PCP and relatively insensitive to the chemical environment [39]. To have a clear picture of the mass-dependence of the effective potential in the EKS equations a comparative study was done on $XCN$ ($X = \mu^+, H, D, T$) species, where the carbon and nitrogen nuclei were considered as clamped point charges while $X$ was treated as a quantum particle. Figure 1 offers the final optimized results of the EKS calculations revealing that the effective potential associated to the heavier particle has less deviations from the point charge coulombic potential far from the banquet center while its one-particle density is more localized. These observations are in line with the expectation that a massive quantum particle behaves more like a clamped point charge than a lighter one. To have a quantified picture of these variations the effective potentials in equations (5) and (7) are rewritten using the known mass-dependence of $\alpha$ from the isotropic harmonic oscillator model [94]:

$$v_{eff}^{e\mu}(\vec{r}_i) = -\frac{1}{|\vec{r}_i - \vec{R}_c|} erf\left[\sqrt[4]{km}\left|\vec{r}_i - \vec{R}_c\right|\right]$$

$$U\left(k, m, \vec{R}_c, \{\vec{R}_\beta\}\right) = \left(\frac{3}{4}\sqrt{\frac{k}{m}}\right) + \sum_\beta^q \frac{Z_\beta}{|\vec{R}_\beta - \vec{R}_c|} erf\left[\sqrt[4]{km}\left|\vec{R}_\beta - \vec{R}_c\right|\right] \quad (11)$$

In these potentials $k$ stands for the force constant which is mainly characteristic of the environment and theoretically, independent from the mass. Thus, the model is to be taken seriously only when slight variation of this effective force constant is seen upon isotopic substitution in the EKS calculations. To test the reliability of the model, Figure 1 compares the mass-dependence of the optimized $\alpha$ and $k$ values demonstrating that the latter is indeed much less sensitive to the mass variations, and may even be treated as a constant in the case of the hydrogen isotopes. This observation reveals that equations (11) are reliable for modeling the intrinsic mass-dependence of the effective potential while the smaller variations of the potentials induced by the variations of the environment are all absorbed in the effective force constant. If ab initio EKS calculations demonstrate that the variations of the effective force constant in a certain set of muonic species are also small, then it is legitimate to introduce a "mean" $\alpha$ and corresponding mean effective potential for the corresponding set. The introduction of the mean effective potentials particularly bypasses the costly non-linear optimization of $\alpha$ for each species separately.

Seven small organic molecules: diazene, acetylene, methenamine, hydrogen cyanide, formamide, fomaldehyde, and ethylene, where their atom types and bonds are typical to many organic molecules used in the experimental μSR studies [1,3], were selected as the representative set. Since in the corresponding experiments muonium atom ($\mu^+$ plus an electron) is attached to organic molecules, we have also attached muonium atom to these seven molecules and the resulting set of open-shell muonic species was employed in order to evaluate the numerical values of $\alpha$. Practically, to have an initial geometry, a hydrogen atom with a clamped proton was first attached to the molecules of the representative set and from various possible conformers the lower energy minima were extracted after the geometry optimization at B3LYP/6-311++g(d,p) level. In next step, a muonium atom replaced the hydrogen atom, i.e. a banquet atom and corresponding electronic [4s1p] basis functions were added instead of hydrogen atom. Subsequently, the position of the banquet atoms, $\alpha$ and the exponents of [4s1p] electronic basis set were simultionsly optimized while the geometry of the remaining nuclei was held fixed. The resulting eleven low-energy muoniated structures are depicted in Figure 2 while the corresponding final optimized geometries have been given in the supporting information. Table 1 offers the optimized $\alpha$ values at both EKS-UB3LYP/[6-311++g(d,p)/4s1p] and EKS-ROB3LYP/[6-311++g(d,p)/4s1p] levels of calculations; the resulting optimized $\alpha$ values are distributed narrowly, $\alpha_{EKS} = 6.05 \pm 0.1$, and insensitive to the U or the RO versions of the EKS calculations. This is not far from the mean $\alpha$ derived from the EHF calculations in the previous study on a completely different set of closed-shell muonic species [39], $\alpha_{EHF} = 5.75$, and confirms that the idea of the mean effective potential is reliable enough to be used in the EKS calculations. On the other hand, in the previous EHF study the mean exponents of the basis functions in the extended muonic [2s2p2d] basis set were narrowly distributed around the mean: $0.8\alpha_{EHF} - 1.4\alpha_{EHF}$ [39]. In present study the exponents of [5s5p] muonic basis set were also scaled around the mean: $0.5\alpha_{EKS}$, $0.75\alpha_{EKS}$, $\alpha_{EKS}$, $1.5\alpha_{EKS}$, $2.0\alpha_{EKS}$. Then, they are used in the construction of corresponding extended effective potential, which in contrast to the simple effective potentials in equations (5) and (7), includes the anharmonicity and the anisotropy of $\mu^+$'s vibrations (for a thorough discussion see [39]). Hence, only the linear coefficients in the extended effective potential were optimized during the EKS calculations (see the supporting information of [39] for details of the procedure). The exponents of the electronic [4s1p] basis set were also optimized both at EKS-UB3LYP/[6-311++g(d,p)/4s1p] and EKS-ROB3LYP/[6-311++g(d,p)/4s1p] levels and the final results have been gathered in Table 2. Once again, it is evident from this table that the optimized exponents are relatively insensitive to chemical environment and the mean values are proper representatives for the optimized exponents. Therefore, the mean values derived at the EKS-UB3LYP/[6-311++g(d,p)/4s1p] level will be used in the rest of this paper. The computed mean exponents are comparable to the mean exponents derived in the previous study using the EHF equations: $\alpha_s = 4.21, 1.20, 0.37, 0.12, \alpha_p = 0.58$ [39], and are insensitive to the U and the RO versions of the EKS calculations as well.

### B. Muoniated ferrocenyl radicals

The idea of adding muonium atoms to organic molecules is not new [1], by the way, more recently muonium atoms have been attached also to carbenes, their organosilicon analogs and organometallic molecules [95,96]. One of the interesting organometallic targets considered both experimentally and computationally is the iconic ferrocene molecule though the corresponding experimental μSR spectrum of its muoniated radical is not yet conclusively assigned [97-99]. Taking the size of ferrocene molecule, all previous computational studies employed DFT to model this system where instead of the muonium atom, a hydrogen atom with a clamped nucleus has been added to ferrocene [97,99]. In this section the EKS-UB3LYP/[6-311++g(d,p)/4s1p] method is used in conjunction with the extended effective potential developed in the previous section in order to study the muoniated ferrocenyl radicals.

In order to start the calculations, we reoptimized the ferrocenyl radical structures reported by McKenzie at UB3LYP/6-311++g(d,p) level [99]. The four optimized structures include ferrocenyl radicals after hydrogen addition to cyclopentadienyl ring or the iron atom while considering the relative configuration of the cyclopentadienyl (Cp) rings that could be staggered or eclipsed [99]. In the next step, the hydrogen atom was eliminated from the structures and a banquet atom with a [4s1p] basis set was added instead. In the case of radicals originating from adding hydrogen atom to the Cp ring, banquet atom could be placed at exo, i.e. with the farthest distance to the iron center, or endo, i.e. with the closest distance to the iron center, positions. Therefore, six distinct muoniated structures were prepared and the EKS-UB3LYP/[6-311++g(d,p)/4s1p] method along with the geometry optimization were applied. Figure 3 depicts the final optimized structures and the used nomenclature while supporting information contains the optimized coordinates. In the case of exo-Cp-eclipsed, endo-Cp-eclipsed and Fe-staggered structures, the full geomtry optimization did not yield stable structures thus they were derived by imposing a plane of symmetry as a contriant. Indeed, the hydorgenic analogs of these structures are saddle points on the corresponding energy hypersurface as reproted by McKenzie and independently confirmed in the present study [99]. Figure 3 also contains the "mean "distance between $\mu^+$ and neighboring nucleus; it is important to stress that this mean distance is distinct from the distance between banquet center and neighboring nucleus and is the expectation value of $\mu^+$'s position operator where the neighboring nucleus is at the center of the coordinate system. The mean C-$\mu$ distances, 1.16 - 1.18 Å, if compared to usual C-H distances, 1.08 - 1.10 Å, are quite longer revealing the nonnegligible elongations of the bond lengths upon isotopic substitution. This trend is completely absent in the conventional ab initio calculations based on using a clamped hydrogen atom to model muonium addition to molecular systems. This observation is in line with one of our previous studies where substitution of $\mu^+$ with one of the hydrogen atoms of malonaldehyde varied the conformation drastically [37]. All observed traits point to the fact that even at the structural level the EKS results contain novel information that are absent if the clamped nucleus model be used instead. Table 3 includes the total and relative energies for the considered species revealing that exo-Cp-staggered radical is the most stable species in line with McKenzie's results [99]. As stressed before, (exo/endo)-Cp-ecllipesed radicals are not stable structures, and threfore, the next stable structure is the exo-Cp-staggered radical, which is only ~3 kJ/mol less stable than the correspodning exo radical whereas the Fe-ecplised radical is ~53 kJ/mol less table than the exo. Thus, it seems reasonable to assume that only the Cp-staggered conformer has a major contribution in the muoniation in gas phase though as disucssed in detail by McKenzie [99], in the solid-state uncertainties emerge due to the crystal packing effects and possible structural deformations of ferreocne molecule itself.

## V. Conclusion

The present study demonstrates that the basic idea of the effective theory, proposed recently [38,39], is extendable both theoretically and computationally to the DFT of muonic systems. Nevertheless, this is just a primary result and it is desirable to include ee correlation at the wavefunction-based levels of the effective theory as well as devising the effective version of the NEO-DFT(ee+eµ), both will be discussed in subsequent reports. In the meantime, the EKS equations provide a framework to start studying the structural and energetic aspects of large muonic systems though without including eµ correlation the quantitative prediction of the µSR spectrum remains yet elusive. The EKS equations may also yield the required information, e.g. one-particle densities, for "atoms in molecules" study of muonic species as considered in some previous reports [34-37]. This path is also now under scrutiny in our laboratory and the results will be discussed in future communications.

### Conflicts of interest

There are no conflicts of interest to declare.

# Acknowledgments

The authors are grateful to Cina Foroutan-Nejad for detailed reading of this paper.

Tables:

Table 1- The optimized $\alpha$ values (in atomic units) derived at EKS-UB3LYP/[6-311++g(d,p)/4s1p] and EKS-ROB3LYP/[6-311++g(d,p)/4s1p] levels for the eleven representative set of the muoniated species depicted in Figure 2.

| backbone | U | RO |
|---|---|---|
| **Acetylene** | 6.10 | 6.10 |
| **Diazene** | 6.10 | 6.10 |
| **Ethylene** | 6.16 | 6.16 |
| **C-Formaldehyde** | 5.98 | 5.99 |
| **O-Formaldehyde** | 5.98 | 5.98 |
| **C-Formamide** | 6.07 | 6.07 |
| **O-Formamide** | 5.91 | 5.91 |
| **C-Hydrogen cyanide** | 6.02 | 6.03 |
| **N-Hydrogen cyanide** | 5.95 | 5.95 |
| **C-Methenamine** | 6.15 | 6.16 |
| **N-Methenamine** | 6.14 | 6.15 |
| **mean** | 6.05 | 6.05 |

Table 2- The optimized exponents (in atomic units) of [4s1p] electronic basis set derived at EKS-UB3LYP/[6-311++g(d,p)/4s1p] and EKS-ROB3LYP/[6-311++g(d,p)/4s1p] levels for the eleven representative set of the muoniated species depicted in Figure 2.

| U | S | S | S | S | P |
|---|---|---|---|---|---|
| **Acetylene** | 3.90 | 1.02 | 0.35 | 0.13 | 0.81 |
| **Diazene** | 3.87 | 0.98 | 0.31 | 0.09 | 0.79 |
| **Ethylene** | 3.89 | 0.99 | 0.31 | 0.11 | 0.87 |
| **C-Formaldehyde** | 3.76 | 0.96 | 0.29 | 0.09 | 0.80 |
| **O-Formaldehyde** | 4.31 | 1.17 | 0.38 | 0.11 | 0.68 |
| **C-Formamide** | 3.85 | 0.97 | 0.30 | 0.09 | 0.97 |
| **O-Formamide** | 4.24 | 1.17 | 0.39 | 0.12 | 0.71 |
| **C-Hydrogen cyanide** | 3.51 | 0.86 | 0.27 | 0.08 | 0.88 |
| **N-Hydrogen cyanide** | 3.91 | 1.02 | 0.33 | 0.09 | 0.70 |
| **C-Methenamine** | 3.75 | 0.94 | 0.29 | 0.09 | 0.88 |
| **N-Methenamine** | 4.19 | 1.09 | 0.36 | 0.10 | 0.77 |
| **mean** | 3.92 | 1.01 | 0.33 | 0.10 | 0.81 |

| RO | S | S | S | S | P |
|---|---|---|---|---|---|
| **Acetylene** | 3.72 | 0.94 | 0.31 | 0.11 | 0.82 |
| **Diazene** | 3.78 | 0.96 | 0.31 | 0.08 | 0.79 |
| **Ethylene** | 3.73 | 0.94 | 0.29 | 0.09 | 0.88 |
| **C-Formaldehyde** | 3.70 | 0.94 | 0.29 | 0.09 | 0.81 |
| **O-Formaldehyde** | 4.34 | 1.18 | 0.39 | 0.11 | 0.68 |
| **C-Formamide** | 3.77 | 0.95 | 0.30 | 0.09 | 0.96 |
| **O-Formamide** | 4.30 | 1.18 | 0.39 | 0.12 | 0.71 |
| **C-Hydrogen cyanide** | 3.43 | 0.84 | 0.26 | 0.07 | 0.88 |
| **N-Hydrogen cyanide** | 3.86 | 1.00 | 0.33 | 0.09 | 0.70 |
| **C-Methenamine** | 3.67 | 0.92 | 0.28 | 0.08 | 0.88 |
| **N-Methenamine** | 4.10 | 1.07 | 0.35 | 0.10 | 0.77 |
| **mean** | 3.85 | 0.99 | 0.32 | 0.09 | 0.81 |

Table 3- The total energies (E) in Hartrees, and the energy difference (ΔE) relative to exo-Cp-staggered species in kJ/mol, all computed at the EKS-UB3LYP/[6-311++g(d,p)/4s1p] level for the six muoniated ferrocenyl radicals depicted in Figure 3.

| Radicals | E | ΔE |
|---|---|---|
| exo-Cp-eclipsed | -1651.35270 | 2.4 |
| endo-Cp-eclipsed | -1651.35153 | 5.4 |
| Fe-eclipsed | -1651.33346 | 52.9 |
| exo-Cp-staggered | -1651.35360 | 0.0 |
| endo-Cp-staggered | -1651.35246 | 3.0 |
| Fe-staggered | -1651.33227 | 56.0 |

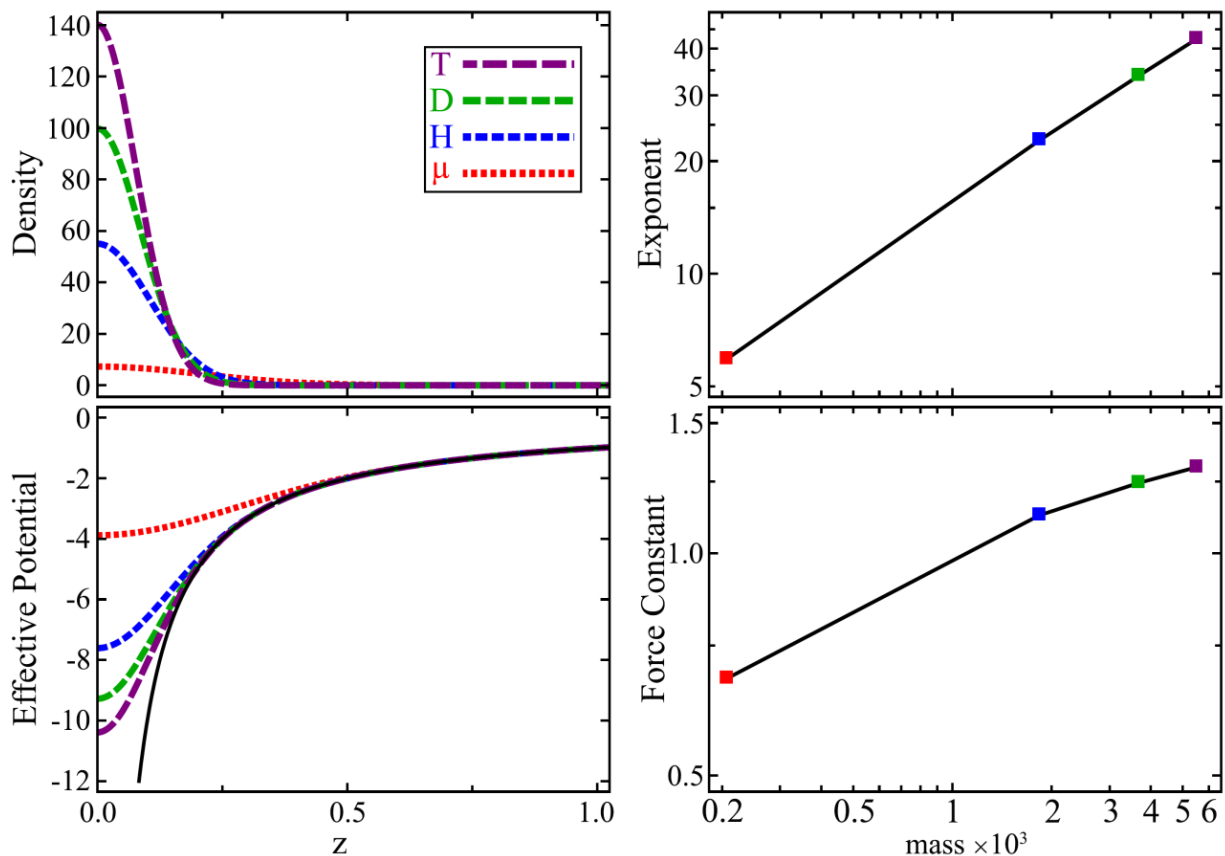

Figure 1- The one-particle density and the optimized effective potential (see equation (7) in the text) of $XCN$ ($X = \mu, H, D, T$) series of species obtained from the EKS-B3LYP/[6-311++g(d,p)/4s1p] calculations. The black line in the effective potential panel is the point charge coulombic potential (proportional to the inverse of distance). The optimized parameter of the effective potential, i.e. the exponent of the s-type GTF, and the corresponding effective force constants are: $\alpha_\mu = 5.91$, $\alpha_H = 22.73$, $\alpha_D = 33.80$, $\alpha_T = 42.39$, and $k_\mu = 0.68$, $k_H = 1.13$, $k_D = 1.24$, $k_T = 1.31$, respectively. All quantities are given in atomic units.

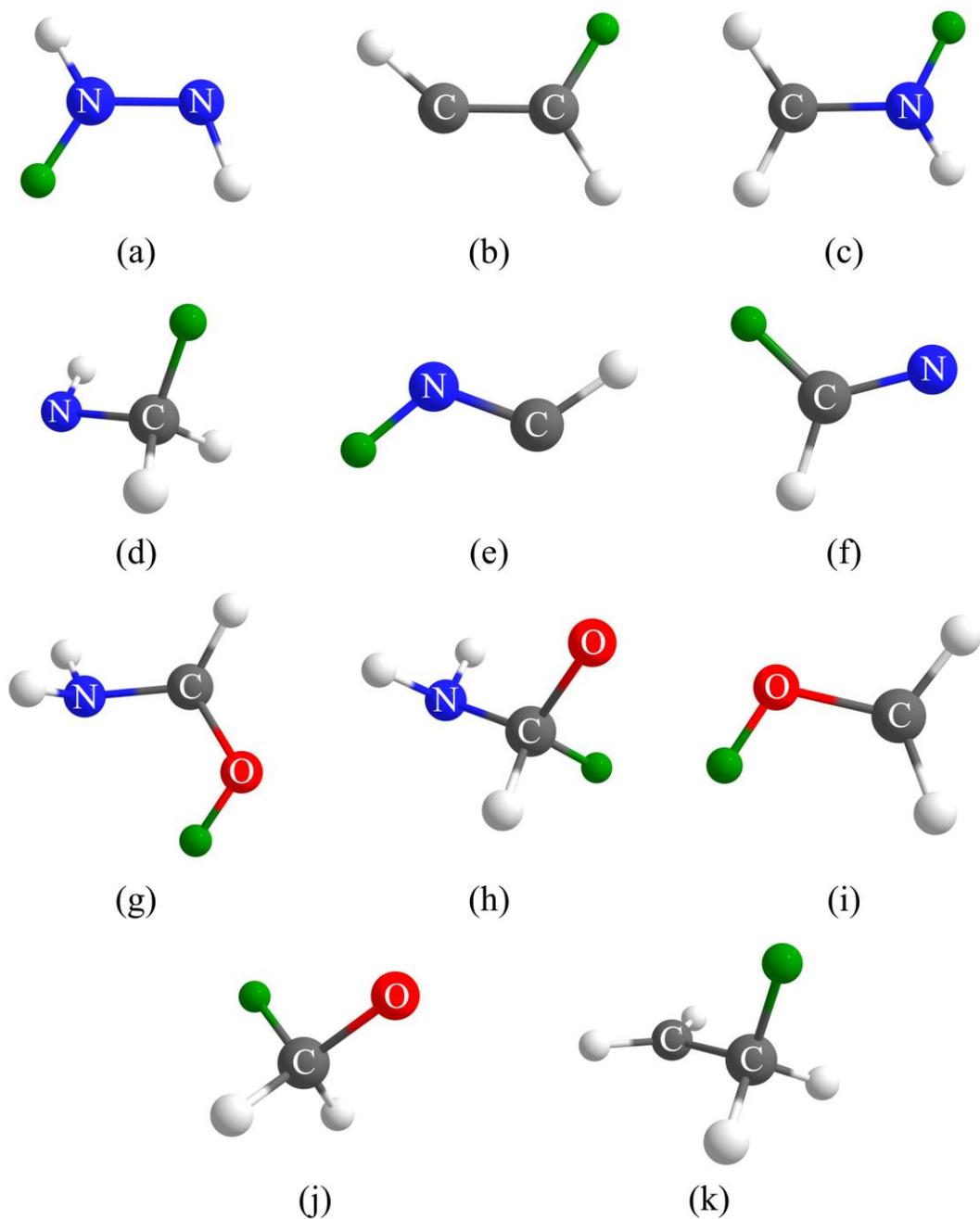

Figure 2- The structures of muoniated (a) diazene, (b) acetylene, (c) N-methenamine, (d) C-methenamine, (e) N-hydrogen cyanide, (f) C-hydrogen cyanide, (g) O-formamide, (h) C-formamide, (i) O-fomaldehyde, (j) C-fomaldehyde, and (k) ethylene adducts (N, C and O symbols are used to descriminate muon's attachment site). Blue, grey, red, white and green spheres indicate the locations of nitrogen, carbon, oxygen, hydrogen and muonium banquet atoms, respectively.

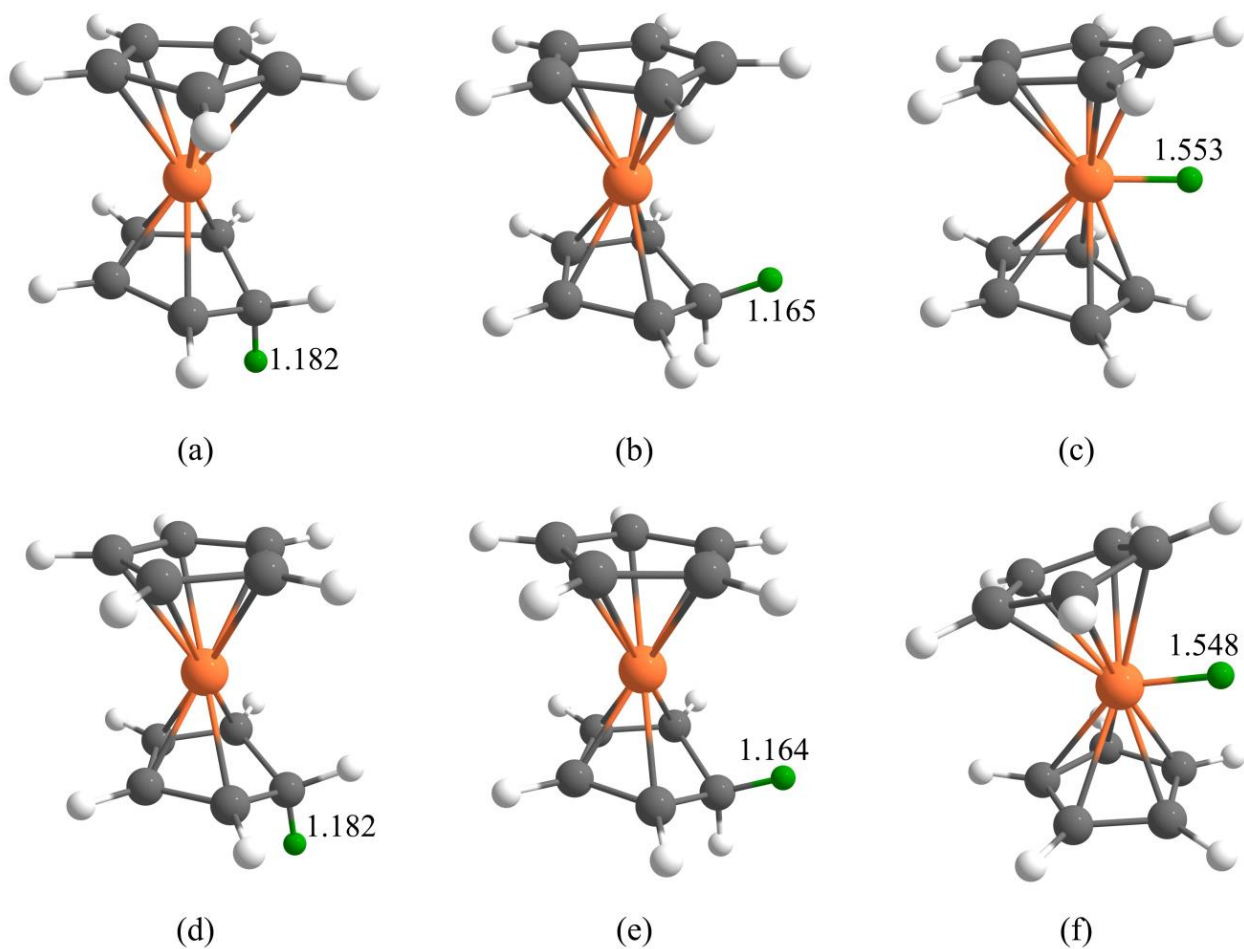

Figure 3- The equilibriumm structures of ferrocene muoniated adducts (a) exo-Cp-eclipsed, (b) endo-Cp-eclipsed, (c) Fe-eclipsed, (d) exo-Cp-staggered, (e) endo-Cp-staggered and (f) Fe-staggered. The mean inter-nuclear distance between muonium (banquet) atom and its binding site (neigboring nucleus) for each adduct has been given over the corresponding bond (in angstroms). Orange, grey, white and green spheres indicate the locations of iron, carbon, hydrogen and muonium (banquet) atoms, respectively.

# Supporting Information

**Effective electronic-only Kohn-Sham equations for the muonic molecules**

Milad Rayka[1], Mohammad Goli[2,*] and Shant Shahbazian[1,*]

[1] *Department of Physics and Department of Physical and Computational Chemistry, Shahid Beheshti University, G. C., Evin, Tehran, Iran, 19839, P.O. Box 19395-4716.*

[2] *School of Nano Science, Institute for Research in Fundamental Sciences (IPM), Tehran 19395-5531, Iran*

E-mails:

Mohammad Goli : mgoli2019@gmail.com

Shant Shahbazian: sh_shahbazian@sbu.ac.ir

[*] Corresponding authors

# Table of contents

| Mu-acetylene | | | | |
|---|---|---|---|---|
| Atom type | Nuclear charge | Coordinates (Angstrom) | | |
| | | X | Y | Z |
| C | 6 | 0.04801 | -0.58616 | 0.00000 |
| Mu | 1 | -0.94683 | -1.20419 | 0.00000 |
| H | 1 | 0.96878 | -1.16584 | 0.00000 |
| C | 6 | 0.04801 | 0.71926 | 0.00000 |
| H | 1 | -0.66429 | 1.53126 | 0.00000 |

| Mu-diazene | | | | |
|---|---|---|---|---|
| Atom type | Nuclear charge | Coordinates (Angstrom) | | |
| | | X | Y | Z |
| N | 7 | 0.74076 | -0.15096 | 0.02319 |
| H | 1 | 1.15164 | 0.78554 | -0.03051 |
| N | 7 | -0.59555 | 0.02444 | -0.06746 |
| H | 1 | -1.13599 | -0.80167 | 0.14495 |
| Mu | 1 | -1.07993 | 0.96698 | 0.17820 |

| Mu-ethylene | | | | |
|---|---|---|---|---|
| Atom type | Nuclear charge | Coordinates (Angstrom) | | |
| | | X | Y | Z |
| C | 6 | -0.69369 | 0.00000 | -0.00200 |
| H | 1 | -1.10645 | -0.88654 | -0.49274 |
| H | 1 | -1.10645 | 0.88653 | -0.49274 |
| Mu | 1 | -1.12735 | 0.00000 | 1.09631 |
| C | 6 | 0.79390 | 0.00000 | -0.01808 |
| H | 1 | 1.35196 | 0.92631 | 0.03997 |
| H | 1 | 1.35196 | -0.92631 | 0.03997 |

| Mu-C-formaldehyde | | | | |
|---|---|---|---|---|
| Atom type | Nuclear charge | Coordinates (Angstrom) | | |
| | | X | Y | Z |
| O | 8 | -0.79134 | 0.00000 | 0.00732 |
| C | 6 | 0.57435 | 0.00000 | 0.01424 |
| H | 1 | 1.00646 | 0.91035 | 0.45537 |
| H | 1 | 1.00646 | -0.91034 | 0.45537 |
| Mu | 1 | 0.89218 | 0.00000 | -1.13575 |

| Mu-O-formaldehyde | | | | |
|---|---|---|---|---|
| Atom type | Nuclear charge | Coordinates (Angstrom) | | |
| | | X | Y | Z |
| C | 6 | -0.68511 | -0.02779 | 0.05962 |
| H | 1 | -1.11960 | -0.99600 | -0.15903 |
| H | 1 | -1.23591 | 0.88841 | -0.09062 |
| O | 8 | 0.67048 | 0.12557 | -0.02145 |
| Mu | 1 | 1.15299 | -0.78134 | 0.08705 |

| Mu-C-formamide | | | | |
|---|---|---|---|---|
| Atom type | Nuclear charge | Coordinates (Angstrom) | | |
| | | X | Y | Z |
| N | 7 | 1.19216 | -0.10187 | 0.00000 |
| H | 1 | 1.35101 | -0.67568 | 0.82140 |
| H | 1 | 1.35101 | -0.67567 | -0.82141 |
| O | 8 | -1.21663 | -0.37191 | 0.00000 |
| C | 6 | -0.13539 | 0.46225 | 0.00000 |
| Mu | 1 | -0.27518 | 1.18353 | -0.93586 |
| H | 1 | -0.25088 | 1.13307 | 0.87248 |

| Mu-O-formamide | | | | |
|---|---|---|---|---|
| Atom type | Nuclear charge | Coordinates (Angstrom) | | |
| | | X | Y | Z |
| N | 7 | -1.12306 | -0.25625 | -0.04393 |
| H | 1 | -1.32328 | -0.69501 | 0.85732 |
| H | 1 | -1.94472 | 0.22766 | -0.38081 |
| C | 6 | 0.04227 | 0.52813 | -0.08608 |
| H | 1 | 0.11354 | 1.51626 | 0.36304 |
| O | 8 | 1.21051 | -0.17384 | 0.04520 |
| Mu | 1 | 1.06747 | -1.10120 | -0.40316 |

| Mu-C-HCN | | | | |
|---|---|---|---|---|
| Atom type | Nuclear charge | Coordinates (Angstrom) | | |
| | | X | Y | Z |
| C | 6 | 0.00000 | 0.50241 | 0.00000 |
| H | 1 | 0.93693 | 1.07817 | 0.00000 |
| Mu | 1 | -1.00799 | 1.11732 | 0.00000 |
| N | 7 | 0.00000 | -0.73868 | 0.00000 |

| Mu-N-HCN | | | | |
|---|---|---|---|---|
| Atom type | Nuclear charge | Coordinates (Angstrom) | | |
| | | X | Y | Z |
| C | 6 | -0.00030 | 0.64744 | 0.00000 |
| H | 1 | 0.89589 | 1.28025 | 0.00000 |
| N | 7 | -0.00030 | -0.58320 | 0.00000 |
| Mu | 1 | -0.95653 | -1.12482 | 0.00000 |

| Mu-C-methenamine | | | | |
|---|---|---|---|---|
| Atom type | Nuclear charge | Coordinates (Angstrom) | | |
| | | X | Y | Z |
| N | 7 | -0.80316 | -0.15310 | 0.00000 |
| H | 1 | -1.21222 | 0.78947 | 0.00000 |
| C | 6 | 0.62833 | 0.01211 | 0.00000 |
| H | 1 | 1.12743 | -0.95854 | 0.00000 |
| H | 1 | 0.96848 | 0.58405 | -0.87858 |
| Mu | 1 | 1.00018 | 0.61822 | 0.94259 |

| Mu-N-methenamine | | | | |
|---|---|---|---|---|
| Atom type | Nuclear charge | Coordinates (Angstrom) | | |
| | | X | Y | Z |
| C | 6 | -0.72897 | 0.00000 | 0.07812 |
| H | 1 | -1.24362 | -0.93097 | -0.11838 |
| H | 1 | -1.24363 | 0.93097 | -0.11839 |
| N | 7 | 0.65504 | 0.00000 | -0.09205 |
| H | 1 | 1.13790 | 0.83602 | 0.20621 |
| Mu | 1 | 1.17885 | -0.89694 | 0.20848 |

| exo-Cp-staggered | | | | |
|---|---|---|---|---|
| Atom | Charge | Coordinates (Angstrom) | | |
| | | X | Y | Z |
| C | 6 | 4.58331 | 1.34874 | 0.00087 |
| C | 6 | 4.28455 | 0.57184 | -1.14566 |
| C | 6 | 4.28319 | 0.56404 | 1.14185 |
| C | 6 | 3.86885 | -0.72852 | -0.71811 |
| C | 6 | 3.86834 | -0.73332 | 0.70525 |
| H | 1 | 4.90815 | 2.37867 | 0.00460 |
| H | 1 | 4.37397 | 0.89710 | -2.17191 |
| H | 1 | 4.37095 | 0.88270 | 2.17031 |
| H | 1 | 3.62205 | -1.55676 | -1.36540 |
| H | 1 | 3.62040 | -1.56562 | 1.34689 |
| Fe | 26 | 2.43015 | 0.64833 | -0.00339 |
| C | 6 | 1.09284 | 2.00836 | -0.72137 |
| C | 6 | 1.09351 | 2.01294 | 0.70742 |
| C | 6 | 0.71950 | 0.70493 | -1.14651 |
| C | 6 | 0.72062 | 0.71223 | 1.14128 |
| C | 6 | 0.00000 | 0.00000 | 0.00000 |
| H | 1 | 1.39915 | 2.82811 | -1.35629 |
| H | 1 | 1.40042 | 2.83661 | 1.33696 |
| H | 1 | 0.62301 | 0.41484 | -2.18441 |
| H | 1 | 0.62498 | 0.42872 | 2.18108 |
| Mu | 1 | -1.11732 | 0.16641 | -0.00004 |
| H | 1 | 0.15567 | -1.08243 | 0.00327 |

| endo-Cp-staggered | | | | |
|---|---|---|---|---|
| Atom | Charge | Coordinates (Angstrom) | | |
| | | X | Y | Z |
| C | 6 | 4.58959 | 1.34689 | -0.00093 |
| C | 6 | 4.28926 | 0.56055 | -1.14067 |
| C | 6 | 4.29194 | 0.57174 | 1.14694 |
| C | 6 | 3.87466 | -0.73605 | -0.70195 |
| C | 6 | 3.87574 | -0.72916 | 0.72149 |
| H | 1 | 4.91411 | 2.37691 | -0.00633 |
| H | 1 | 4.37660 | 0.87777 | -2.16960 |
| H | 1 | 4.38222 | 0.89845 | 2.17265 |
| H | 1 | 3.62743 | -1.56946 | -1.34247 |
| H | 1 | 3.63051 | -1.55672 | 1.37025 |
| Fe | 26 | 2.43573 | 0.64653 | 0.00546 |
| C | 6 | 1.10122 | 2.01238 | -0.70383 |
| C | 6 | 1.09970 | 2.00581 | 0.72470 |
| C | 6 | 0.72695 | 0.71206 | -1.13968 |
| C | 6 | 0.72446 | 0.70162 | 1.14773 |
| C | 6 | 0.00000 | 0.00000 | 0.00000 |
| H | 1 | 1.41047 | 2.83640 | -1.33192 |
| H | 1 | 1.40765 | 2.82386 | 1.36116 |
| H | 1 | 0.63355 | 0.43088 | -2.18035 |
| H | 1 | 0.62905 | 0.41062 | 2.18552 |
| H | 1 | -1.09212 | 0.17558 | -0.00036 |
| Mu | 1 | 0.14839 | -1.09955 | -0.00487 |

| exo-Cp-eclipsed | | | | |
|---|---|---|---|---|
| Atom | Charge | Coordinates (Angstrom) | | |
| | | X | Y | Z |
| C | 6 | 4.54825 | -0.70758 | -0.50978 |
| C | 6 | 4.54736 | 0.70860 | -0.50968 |
| C | 6 | 3.99679 | -1.14808 | 0.73334 |
| C | 6 | 3.99536 | 1.14831 | 0.73347 |
| C | 6 | 3.68567 | -0.00011 | 1.50973 |
| H | 1 | 4.89490 | -1.34612 | -1.30860 |
| H | 1 | 4.89328 | 1.34772 | -1.30835 |
| H | 1 | 3.86371 | -2.17556 | 1.03752 |
| H | 1 | 3.86113 | 2.17549 | 1.03806 |
| H | 1 | 3.25212 | -0.00020 | 2.49933 |
| Fe | 26 | 2.48866 | -0.00043 | -0.28915 |
| C | 6 | 1.37178 | -0.71599 | -1.83149 |
| C | 6 | 1.37217 | 0.71435 | -1.83182 |
| C | 6 | 0.80926 | -1.14610 | -0.60075 |
| C | 6 | 0.81001 | 1.14541 | -0.60117 |
| C | 6 | 0.00000 | 0.00000 | 0.00000 |
| H | 1 | 1.79325 | -1.34773 | -2.60084 |
| H | 1 | 1.79403 | 1.34533 | -2.60157 |
| H | 1 | 0.66954 | -2.18453 | -0.33126 |
| H | 1 | 0.67083 | 2.18412 | -0.33246 |
| Mu | 1 | -1.08406 | 0.00044 | -0.31808 |
| H | 1 | 0.00219 | -0.00020 | 1.09405 |

| endo-Cp-eclipsed | | | | |
|---|---|---|---|---|
| Atom | Charge | Coordinates (Angstrom) | | |
| | | X | Y | Z |
| C | 6 | 4.55382 | -0.70764 | -0.50563 |
| C | 6 | 4.55468 | 0.70865 | -0.50594 |
| C | 6 | 4.00353 | -1.14713 | 0.73820 |
| C | 6 | 4.00478 | 1.14926 | 0.73777 |
| C | 6 | 3.69465 | 0.00128 | 1.51457 |
| H | 1 | 4.89870 | -1.34701 | -1.30454 |
| H | 1 | 4.90035 | 1.34704 | -1.30529 |
| H | 1 | 3.86964 | -2.17414 | 1.04350 |
| H | 1 | 3.87219 | 2.17686 | 1.04183 |
| H | 1 | 3.26296 | 0.00067 | 2.50501 |
| Fe | 26 | 2.49499 | 0.00159 | -0.28506 |
| C | 6 | 1.38124 | -0.71424 | -1.82773 |
| C | 6 | 1.38021 | 0.71588 | -1.82791 |
| C | 6 | 0.81626 | -1.14493 | -0.59762 |
| C | 6 | 0.81450 | 1.14602 | -0.59798 |
| C | 6 | 0.00000 | 0.00000 | 0.00000 |
| H | 1 | 1.80631 | -1.34529 | -2.59572 |
| H | 1 | 1.80399 | 1.34777 | -2.59596 |
| H | 1 | 0.67950 | -2.18398 | -0.32889 |
| H | 1 | 0.67555 | 2.18486 | -0.32948 |
| H | 1 | -1.05768 | -0.00098 | -0.32393 |
| Mu | 1 | -0.00786 | 0.00067 | 1.11038 |

| Fe-eclipsed | | | | |
|---|---|---|---|---|
| ATOM | ATOMIC | Coordintes (Angstrom) | | |
| | | X | Y | Z |
| C | 6 | 0.70005 | 1.65236 | 1.34099 |
| C | 6 | -0.70591 | 1.65216 | 1.33737 |
| C | 6 | 0.70579 | -1.65227 | 1.33742 |
| C | 6 | -0.70015 | -1.65240 | 1.34104 |
| C | 6 | 1.15075 | 1.83060 | -0.00339 |
| C | 6 | -1.14985 | 1.83073 | -0.00920 |
| C | 6 | 1.14971 | -1.83085 | -0.00917 |
| C | 6 | -1.15088 | -1.83046 | -0.00334 |
| C | 6 | 0.00259 | 2.01276 | -0.82814 |
| C | 6 | -0.00275 | -2.01271 | -0.82811 |
| H | 1 | 1.33558 | 1.49178 | 2.19883 |
| H | 1 | -1.34558 | 1.49151 | 2.19213 |
| H | 1 | 1.34546 | -1.49170 | 2.19219 |
| H | 1 | -1.33566 | -1.49169 | 2.19887 |
| H | 1 | -2.17620 | 1.89495 | -0.33733 |
| H | 1 | 2.17867 | 1.89477 | -0.32663 |
| H | 1 | -2.17880 | -1.89438 | -0.32660 |
| H | 1 | 2.17607 | -1.89527 | -0.33728 |
| H | 1 | 0.00522 | 2.25816 | -1.87873 |
| H | 1 | -0.00537 | -2.25805 | -1.87871 |
| Fe | 26 | 0.00000 | 0.00000 | 0.00000 |
| Mu | 1 | 0.00002 | 0.00002 | -1.52429 |

| Fe-staggered | | | | |
|---|---|---|---|---|
| Atom | Charge | Coordinates (Angstrom) | | |
|  |  | X | Y | Z |
| C | 6 | -1.64609 | 0.71573 | -1.18184 |
| C | 6 | -0.45784 | 1.14392 | -1.83465 |
| C | 6 | 0.24921 | 0.00078 | -2.28317 |
| C | 6 | -0.45613 | -1.14352 | -1.83543 |
| C | 6 | -1.64489 | -0.71775 | -1.18208 |
| H | 1 | -2.42255 | 1.35442 | -0.79080 |
| H | 1 | -0.15841 | 2.17187 | -1.97759 |
| H | 1 | 1.20444 | 0.00104 | -2.78565 |
| H | 1 | -0.15491 | -2.17089 | -1.97872 |
| H | 1 | -2.42050 | -1.35782 | -0.79166 |
| Fe | 26 | 0.00000 | 0.00000 | 0.00000 |
| C | 6 | 0.65487 | 0.00016 | 2.07423 |
| C | 6 | 1.17236 | 1.15047 | 1.41057 |
| C | 6 | 2.09538 | 0.70096 | 0.41616 |
| C | 6 | 2.09430 | -0.70344 | 0.41629 |
| C | 6 | 1.17068 | -1.15121 | 1.41094 |
| H | 1 | 0.00497 | 0.00104 | 2.93531 |
| H | 1 | 0.96449 | 2.17784 | 1.66811 |
| H | 1 | 2.64835 | 1.33730 | -0.25853 |
| H | 1 | 2.64628 | -1.34093 | -0.25811 |
| H | 1 | 0.96103 | -2.17815 | 1.66877 |
| Mu | 1 | -1.15470 | 0.00104 | 0.98778 |